%% file: manuscript.tex
\documentclass[lettersize,journal]{IEEEtran}
\usepackage{amsmath,amsfonts}
\usepackage{algorithmic}
\usepackage{algorithm}
\usepackage{array}
\usepackage[caption=false,font=normalsize,labelfont=sf,textfont=sf]{subfig}
\usepackage{caption}
\usepackage{color}

\usepackage{textcomp}
\usepackage{stfloats}
\usepackage{url}
\usepackage{verbatim}
\usepackage{graphicx}
\usepackage{cite}
\usepackage{multirow}
\newcommand\blfootnote[1]{%
\begingroup
\renewcommand\thefootnote{}\footnote{#1}%
\addtocounter{footnote}{-1}%
\endgroup
}

\title{Reality3DSketch: Rapid 3D Modeling of Objects from Single Freehand Sketches}

\author{Tianrun Chen, Chaotao Ding, Lanyun Zhu, Ying Zang*, Yiyi Liao*, Zejian Li, and Lingyun Sun
\thanks{.}
\thanks{Manuscript received XXX, 2022; revised XXX X, XXXX.}}

\IEEEpubid{}
\begin{document}
\twocolumn[{%
\renewcommand\twocolumn[1][]{#1}%
\maketitle
\begin{center}
    \centering
    \vspace{-1cm}
    \includegraphics[width=\textwidth]{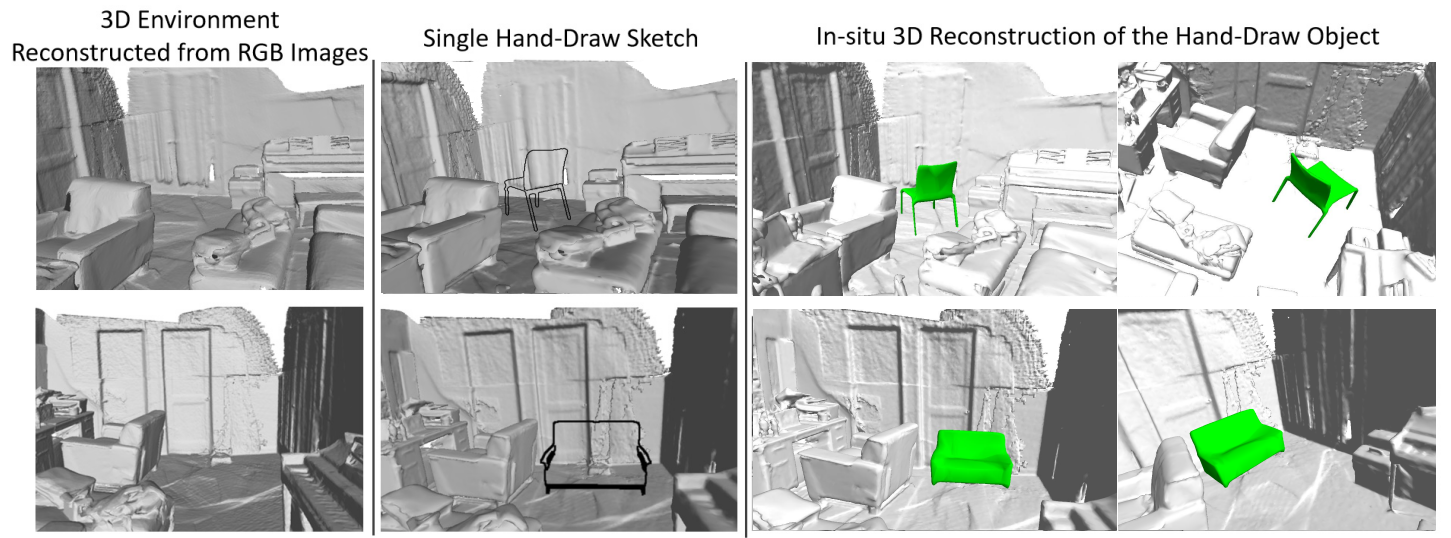}
    \captionof{figure}{Given a collection of RGB images captured by mobile phones, Reality3DSketch  generates a 3D object based on the input of the hand-drawn sketch and places it at the user's desired location. We first obtain a reconstructed 3D environment (first column). Users can then draw a sketch from a specific view (second column), and our algorithm reconstructs black an \textit{in situ} 3D object placed at the user's desired location (third and fourth columns; added objects are highlighted in green for visual clarity).}
\end{center}%
}]

\begin{figure*}
\centering
\includegraphics[width=0.8\textwidth]{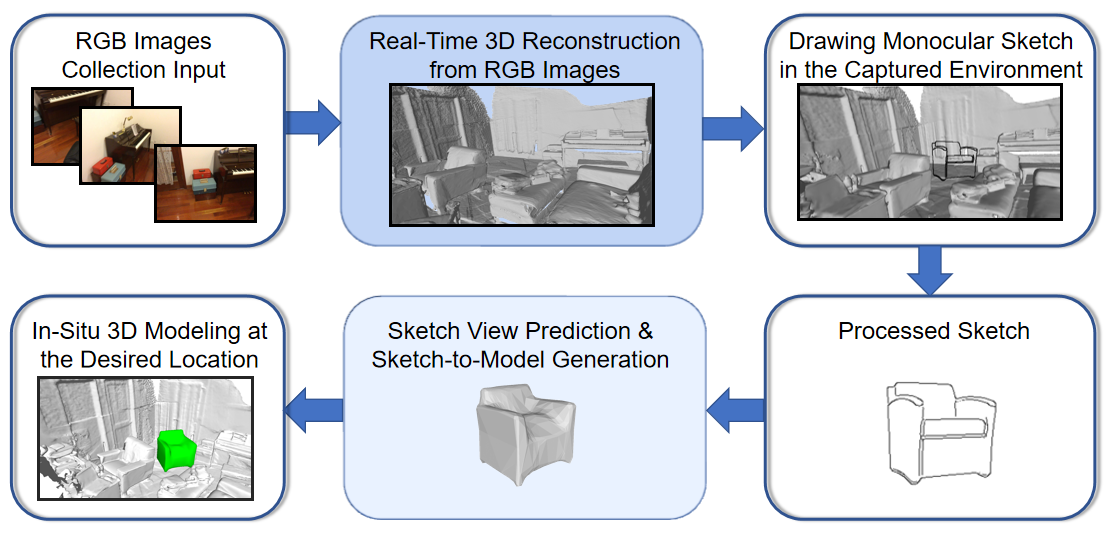}
\caption{\textbf{Overview of Reality3DSketch.} The two blue boxes denote two neural networks. The first network takes the input of the RGB images captured by mobile devices and produces a real-time 3D reconstruction. The reconstructed results are rendered on the screen for users to draw a sketch within the scene. Users draw a sketch at a reasonable location, and the sketch is processed and inputted to another network for view prediction and sketch-to-model generation. Ultimately, an \textit{in situ} 3D model at the user's desired location is obtained.}
\end{figure*}

\blfootnote{Tianrun Chen is with the College of Computer Science and Technology,
Zhejiang University, China, 310027 and KOKONI, Moxin (Huzhou) Technology Co., LTD. (email: tianrun.chen@zju.edu.cn)}
\blfootnote{Zejian Li is with the School of Software Technology, Zhejiang University, China, 310027 (email: zejianlee@zju.edu.cn).}
\blfootnote{Chaotao Ding and Ying Zang are with the School of Information Engineering, Huzhou University, China, 313000 (e-mail: 2021388117@stu.zjhu.edu.cn, 02750@zjhu.edu.cn).}
\blfootnote{Yiyi Liao is with the College of Information Science and Electronic Engineering, Zhejiang University, China, 310027 (email: yiyi.liao@zju.edu.cn).}
\blfootnote{Lanyun Zhu is with the Information Systems Technology
and Design Pillar, Singapore University of Technology and Design, Singapore 487372 (e-mail: lanyun\_zhu@mymail.sutd.edu.sg).}
\blfootnote{Lingyun Sun is with the College of Computer Science and Technology,
Zhejiang University, China, 310027. (email: sunly@zju.edu.cn) }

\blfootnote{This work is an extended version of the conference report of Chen, Tianrun, et al., "Deep3DSketch: 3D Modeling from Free-hand Sketches with View- and Structural-Aware Adversarial Training", at the 2023 IEEE International Conference on Acoustics, Speech and Signal Processing (ICASSP), IEEE, 2023.}
\blfootnote{*Corresponding Author}

\begin{abstract}
The emerging trend of AR/VR places great demands on 3D content. However, most existing software requires expertise and is difficult for novice users to use. In this paper, we aim to create sketch-based modeling tools for user-friendly 3D modeling. We introduce Reality3DSketch with a novel application of an immersive 3D modeling experience, in which a user can capture the surrounding scene using a monocular RGB camera and can draw a single sketch of an object in the real-time reconstructed 3D scene. A 3D object is generated and placed in the desired location, enabled by our novel neural network with the input of a single sketch. Our neural network can predict the pose of a drawing and can turn a single sketch into a 3D model with view and structural awareness, which addresses the challenge of sparse sketch input and view ambiguity. We conducted extensive experiments synthetic and real-world datasets and achieved state-of-the-art (SOTA) results in both sketch view estimation and 3D modeling performance. According to our user study, our method of performing 3D modeling in a scene is $>$5x faster than conventional methods. Users are also more satisfied with the generated 3D model than the results of existing methods. 
\end{abstract}


\section{Introduction}
\IEEEPARstart{T}{he} rapid development of portable displays and the emerging trend of metaverse applications, including AR/VR, bring new possibilities in the digital era -- people can now view massive digital content and even interact with it in a virtual 3D world. This emerging trend calls for a large quantity of versatile and customizable 3D content in the virtual world \cite{chen2022multimedia,wang2020vr}. Much effort has been made to design 3D modeling tools to be simpler and encourage creativity \cite{hurst2013making}.

However, most existing 3D modeling tools cannot meet this demand in the metaverse era. First, they are not friendly to novice users aiming to create customized 3D models. Widely used computer-aided design (CAD) software requires knowledge that involves both sophisticated CAD software commands and strategies, which is required to parse a shape into sequential commands \cite{bhavnani1999strategic, chester2007teaching}; additionally, it is a labor-intensive and time-consuming process \cite{reddy2018development}. Second, most 3D modeling tools isolate the creation process from existing 3D content, and the created model may lack context information and require extra effort to fit it to the VR/AR context \cite{do20103darmodeler}. 

A promising solution to the abovementioned limitations is to utilize sketch-based 3D modeling tools as an alternative to the conventional CAD software suite. As sketching is a natural form of expression for human beings, using sketches as the input to produce 3D models can free users from mastering 3D modeling. In particular, there are 3D sketching tools in the context of VR/AR \cite{arora2018symbiosissketch, deering1995holosketch,kwan2019mobi3dsketch,keefe2001cavepainting,xu2018model}. They allow users to immersively and freely draw 3D curves directly in the air, ensuring that the generated 3D model fits in a 3D world. However, special devices and sufficient expertise are still required to use these tools -- most of the tools require MoCap systems or a motion-tracking stylus to provide precise localization of 3D stokes, and the modeling process is still designed for users with reasonably good drawing skills and not for novice users \cite{kwan2019mobi3dsketch}. The requirement of sufficient expertise and much practice in using existing 3D sketching tools also comes from the depth-perception issue \cite{kwan2019mobi3dsketch,arora2017experimental,machuca2019smart3dguides} -- users have difficulties in localizing the desired drawing position in 3D space, especially in settings where users draw 3D sketches by viewing only a 2D display panel on tablets or mobile phones without depth perception \cite{kwan2019mobi3dsketch}.

To address this challenge, we offer a novel solution, \textbf{Reality3DSketch}, to provide a new paradigm for novice users to create new customized 3D models in a given 3D world. Reality3DSketch is an AI-enabled 3D modeling tool inspired by the recent success of AI-enabled content creation. We first propose a novel generative network that can use only a single-view freehand 2D sketch (in a single plane) as the input to produce a high-fidelity 3D model. This eliminates the need for users to have specialized skills or to provide multiple consistent views of the object, thereby minimizing their effort. Instead of using multiple precise line drawings that require drawing expertise or a step-by-step workflow that requires \textit{strategic knowledge} \cite{ cohen1999interface, deng2020interactive}, Reality3DSketch allows users to draw a single-view sketch \textit{from an arbitrary viewpoint}. The neural network processes the rest, and it is trained to estimate the user-intended viewpoint. Viewpoint estimation can constrain the model generation process to resolve the view ambiguity of the single sketch. The estimated viewpoint can also be used to guide the positioning of the 3D model in a real scene. Since the generated 3D model is aligned in a particular viewpoint as in the dataset, having the viewpoint information can enable the object to be rotated to match the user-desired view angle. 

We design novel modules to enhance the performance of the sketch-to-model process. We disentangle the learning of 3D shapes and viewpoints by random pose sampling (RPS) of the object silhouette, and we input the randomly sampled silhouette to an effective progressive shape discriminator that is aware of the objects' geometric structure via cross-view silhouettes of the 3D model. The network is designed to have both view- and structure-aware properties, aiming to provide high-fidelity 3D modeling that can accurately reflect users' intentions. Furthermore, to offer an immersive creative environment with depth perception, we exploratively propose to have users draw a single-view sketch in a 3D reconstruction of the surrounding environment obtained from images captured by a mobile phone instead of drawing by viewing the raw scenes captured by a monocular RGB camera with a limited field of view. Our experimental results demonstrate that contextual geometric information can be beneficial for users in terms of user experience. With the 3D reconstructed mesh, users can rotate and view the environment from different angles as they sketch. This gives the user a better sense of depth and spatial relationships than with a single flat image. This can be especially useful when designing new components that need to fit into the environment in a specific way. Additionally, the resulting object can be viewed directly from different angles and distances, allowing for easy checking or adjustment immediately after the sketch-to-shape generation process, which is also time-saving for users.

We conducted extensive experiments using both a synthetic dataset and a user-drawn real dataset. The qualitative and quantitative results show the effectiveness of our novel sketch-to-model approach with state-of-the-art (SOTA) performance. As Reality3DSketch is a new 3D interaction paradigm, in which the sketch can precisely define the 6D pose and position of the generated object, we also performed a user study that compared our 3D modeling approach and the conventional manual approach of 3D interaction. The results show that our method is $>$5 times faster than the baseline approach in performing 3D modeling and interaction within a scene. We further performed a user study that compared Reality3DSketch with RGB context input sketch-based 3D modeling. The results showed that involving geometric information led to significantly higher user experience ratings and fewer redo actions. Another user study shows that users were more satisfied with the 3D model generated with our approach, demonstrating the practicality and effectiveness of our novel 3D modeling pipeline. 

Specifically, our contributions are as follows:
\begin{itemize}
\item We propose a novel paradigm, Reality3DSketch, for intuitive and immersive 3D modeling. Table I shows the differences between Reality3DSketch and other pipelines. We consider a case in which users can use their phones to capture the surrounding environment to obtain an accurate 3D reconstruction and draw a single-view sketch at the desired location. The 3D object is generated and put into the virtual environment. Both the 3D reconstruction and the sketch-to-model generation are processed in real time. Our user study shows that our immersive 3D modeling approach is $>$5 times faster than separately modeling and manually positioning the object. 
\item  The sketch-to-model process is realized by a novel neural network we propose. The network is designed to have random pose sampling (RPS) and a progressive shape discriminator (SD) so that it is both view- and structure-aware to ensure high-fidelity model generation.
\item State-of-the-art (SOTA) performance is achieved with both our sketch view estimation result and our 3D modeling result in both synthetic and real datasets. In the user study, users are also more satisfied with the generated 3D model than previous methods.
\end{itemize}

\begin{table*}[ht]
    \centering
    \setlength{\tabcolsep}{1mm}
    \caption{Comparison of Reality3DSketch with Some Other Existing 3D Modeling Approaches}
    \label{table:table0}
    \scalebox{0.8}{
        \begin{tabular}{| c | c | c | c | c | c | c |}  
            \hline
            \multirow{2}{*}{Method} & {Retrieval-Based} & {Rule-Based} & {Deep-Learning} &  \multirow{2}{*}{CAD Modeling} & CAD Modeling + Touch & \multirow{2}{*}{Reality3DSketch} \\ 
             &  Sketch-to-Shape \cite{igarashi2006teddy, li2020sketch2cad, cohen1999interface, shtof2013geosemantic, jorge2003gides++, gingold2009structured, deng2020interactive} &  Sketch-to-Shape \cite{igarashi2006teddy, li2020sketch2cad} &  Sketch-to-Shape \cite{zhang2021sketch2model,guillard2021sketch2mesh} 
 &  & Based Manipulation \cite{martinet2011integrality,goh20193d} & \\ \hline
            Customization Flexibility & & & $\checkmark$ & $\checkmark$ & $\checkmark$ &  $\checkmark$ \\ \hline
            Novice User-Friendly & $\checkmark$ & $\checkmark$ & $\checkmark$ & & & $\checkmark$ \\ \hline
            Context-Aware Manipulation & & & & & $\checkmark$ & $\checkmark$ \\ \hline
            Quick Result Generation & $\checkmark$ & $\checkmark$ & $\checkmark$ & & & $\checkmark$ \\ \hline
        \end{tabular}
    }
\end{table*}

\section{Related Works}

\subsection{2D Sketch-Based 3D Modeling}
Sketch-based 3D modeling has been studied by researchers for decades. Early works mainly focused on drawing 2D sketches on paper or a touch-screen panel, and 3D models were obtained accordingly. Bonnici et al. \cite{bonnici2019sketch} and Olsen et al. \cite{olsen2009sketch} comprehensively reviewed the existing sketch-based 3D modeling approaches. Existing sketch-based 3D modeling methods using 2D sketches as the input can be divided into end-to-end and interactive approaches. The interactive approach requires users with strategic knowledge for sequential step decomposition or specific drawing gestures or annotations \cite{igarashi2006teddy, li2020sketch2cad, cohen1999interface, shtof2013geosemantic, jorge2003gides++, gingold2009structured, deng2020interactive}.  For the end-to-end approach, works that use template primitives or retrieval-based approaches \cite{chen2003visual, wang2015sketch, sangkloy2016sketchy, 2016Interactive, 20183D, huang2016shape, nie2020m, xu2020sketch} can produce  some satisfactory results, but they lack customizability. Some very recent work directly reconstructed a 3D model using deep neural networks and recognized the sketch-based 3D modeling as single-view 3D reconstruction \cite{zhang2021sketch2model, guillard2021sketch2mesh, wang20203d, chen2023deep3dsketch}. However, these methods using an autoencoder structure similar to a single-view 3D reconstruction pipeline can only obtain coarse predictions of the 3D model \cite{zhang2021sketch2model} due to the sparsity and ambiguity of sketches. Specifically, sketches are sparse because they have only a single view, are mostly abstract, lack fine boundary information when drawn by humans, and more critically, lack texture information for depth estimation. This brings considerable uncertainty when learning 3D shapes. In this work, we incorporate the advantage of using 2D sketches, which are intuitive and convenient, and we propose a novel network design to alleviate the sparsity and ambiguity of a single 2D sketch to produce high-fidelity 3D modeling that reflects users' ideas.  

\subsection{Immersive 3D Sketching}
With the emergence of AR/VR, 3D sketching tools were developed. An early attempt at 3D sketching in the context of AR/VR was Holosketch \cite{deering1995holosketch}, which supports creating primitives and freeform tubes and wire geometries in 3D. Later works expanded the possibilities with advances in hardware development \cite{keefe2001cavepainting,arora2018symbiosissketch,xu2018model}. 
There are even commercial tools (e.g., Tilt Brush, GravitySketch, and Quill) available for users to directly draw 3D objects in a virtual environment. With various curve- and surface-fitting techniques, even 3D CAD models can be created \cite{wesche2001freedrawer,chen2003visual}. However, despite the freedom of painting with 3D strokes, 3D sketching and potentially forming 3D models is not a trivial task due to two significant challenges. 

The first challenge is the depth perception issue. Distance underestimation \cite{renner2013perception} and disparities in targeting accuracy between lateral and depth motions are frequently found in 3D sketching systems \cite{machuca2019smart3dguides}. Specifically, in a user study of a mobile-based 3D sketching system creating content in a scene captured by RGB cameras, all users reported difficulty in depth estimation when creating the content \cite{kwan2019mobi3dsketch}. 

Another challenge is the high cognitive and sensorimotor demands of drawing in 3D. 
Wiese et al. \cite{wiese2010investigating} discovered that 3D drawing requires more manual effort and  higher cognitive and sensorimotor demands than 2D drawing, which is due to the requirement for users to control more degrees of freedom (DOFs) during movement (3/6 DOFs instead of 2 DOFs). Arora et al. \cite{arora2017experimental} reported that in pure 3D interactive settings without a physical surface, users are forced to rely solely on eye-hand coordination to control stroke position, which introduces extra challenges for creators. 

In contrast, this work demonstrates an application that combines 2D sketching on a physical surface (mobile device) and 3D scene reconstruction (using a regular mobile device) for the first time. There are no longer heavy cognitive and sensorimotor demands, but the generated 3D model can still be fitted in a real scene for AR/VR applications.

\section{Method}
\subsection{Overview}
The overall pipeline of Reality3DSketch is illustrated in Figure 2. We separated the generation of the surrounding environment and sketch-based object 3D modeling into two steps with separate neural networks. Users use their phones to capture the surrounding environment. The obtained posed images are fed into a real-time reconstruction network that directly reconstructs local surfaces, represented as sparse TSDF volumes. The mesh is extracted and rendered on the user's screen. The user then draws one sketch at a single viewpoint within the scene. The single-view sketch is fed into a sketch-to-model network to obtain a complete 3D model. The sketch-to-model network has a view-prediction network to obtain the predicted viewpoint information. Because the object generated from the sketch-to-model network is aligned as in the dataset, a pose transformation (rotation) is performed to translate the object in canonical space to global coordinates, and the sketch-derived 3D model is placed in the reconstructed scene at the users' desired position and angle.
\begin{figure*}
\centering
\includegraphics[width=0.95\textwidth]{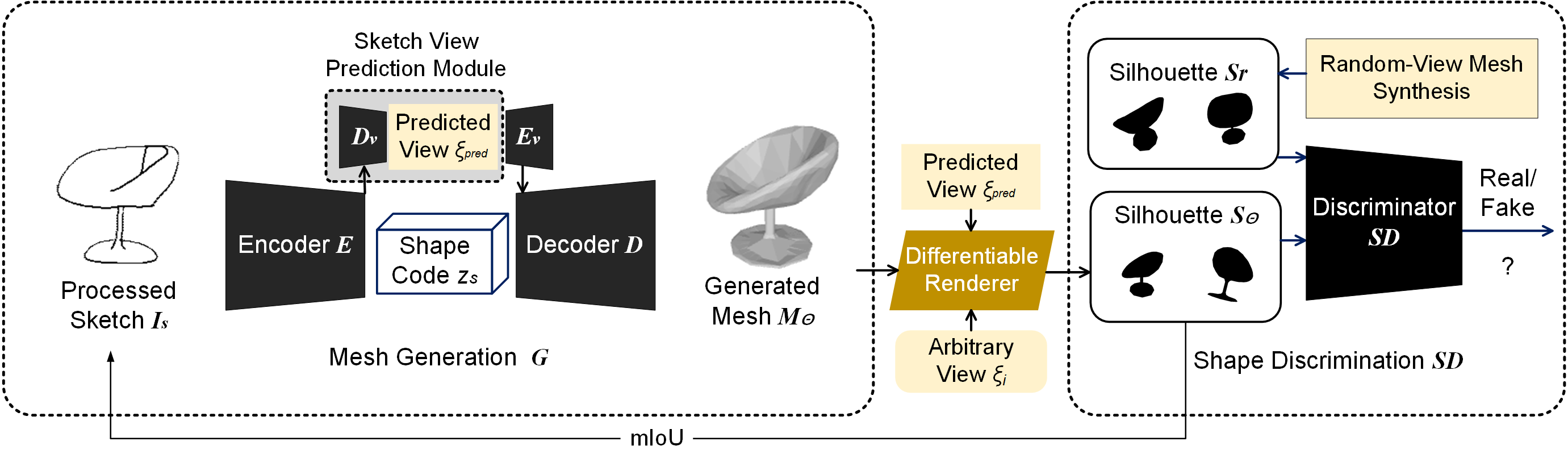}
\caption{\textbf{Pipeline of the Sketch-to-Model Generation Network.} The input sketch is fed into a mesh generation network to produce a generated mesh. The mesh generation network consists of a sketch view prediction module that outputs the predicted viewpoint of the sketch. A shape discriminator is introduced to add extra supervision to produce more realistic models.}

\end{figure*}
\subsection{Preliminaries}
For the 3D reconstruction network, we are given a set of images $\left\{ I_t \right\}$ with the corresponding camera poses $\left \{ \xi_t  \in \mathbb{SE} \left( 3 \right) \right \}$ to obtain a dense 3D mesh reconstruction $M_t$. 

For the sketch-to-model network, the input is a binary sketch $I_s  \in \left \{ 0,1 \right \}^{W\times H}$. We let $I_s \left [ i,j \right ] = 0 $ if it is marked by a pen stroke, and $I_s\left [ i,j \right ] = 1 $ otherwise. The goal of the sketch-to-model network $G$ is to obtain a mesh $M_\Theta =(V_\Theta,  F_\Theta)$, in which $V_\Theta$ and $F_\Theta$ represent the mesh vertices and faces and the silhouette $S_\Theta :\mathbb{R}^3 \rightarrow \{ 0, 1 \} ^{W\times H} $ of $M_\Theta $ best matches the information from the input sketch $I_s$. Compared to NeRF or other 3D representations \cite{zhang2023dyn,zhang2023painting}, the mesh representation of generated shapes offers a seamless integration into reconstructed scenes \cite{dou2022tore,lin2023patch}.

\subsection{View Ambiguity and Sketch View Prediction}
In the sketch-to-model network, we first explicitly learn the viewpoint of the model, which is a fundamental element in positioning the generated 3D model in a scene. We use an encoder $E$ to produce latent code $z_l$ and input it to the viewpoint prediction module, which consists of two fully connected layers $D_v$ to produce the viewpoint estimation $\xi_{pred}$, represented by an Euler angle. The viewpoint prediction module is optimized in a fully supervised manner with the input of the ground-truth viewpoint $\xi_{gt}$, supervised by a viewpoint prediction loss $\mathcal{L}_{v}$, which adopts MSE loss for the predicted and ground-truth poses, defined as:
\begin{align}
\mathcal{L}_{v}=\|\xi_{gt}-\xi_{pred}\|_{2}=\left\|\xi_{gt}-D_{v}\left(z_{l}\right)\right\|_{2}
\end{align}

We also integrate view prediction into the sketch-to-model process, as previous works argue that view ambiguity is a critical issue of sketch-based 3D modeling\cite{zhang2021sketch2model}, which will be illustrated in the subsequent section. 

\subsection{3D Model Generation with View Awareness}
We take a commonly used encoder-decoder structure as the backbone of our sketch-to-model process, as it is a cross-domain prediction task. We use an encoder $E$ to obtain a compressed shape code $z_s$ and a decoder $D$ to manipulate $z_s$ to calculate the vertex offsets of the template mesh and deform it to obtain the output mesh $M_\Theta = D(z_s)$. The silhouette $S_\Theta :\mathbb{R}^3 \rightarrow \{ 0, 1 \} ^{W\times H} $ of $M_\Theta $ should match the input sketch $I_s$. Therefore, we render the silhouette $S_1$ of $M_\Theta $. Specifically, the output viewpoint prediction $\xi_{pred}$ is fed into a differentiable renderer to render a silhouette at the given viewpoint for supervision. We use the mIoU loss $\mathcal{L}_{iou}$ to measure the similarity between the rendered silhouette $S_1$ and the silhouette of the input sketch $S_2$:
\begin{align}
\mathcal{L}_{i o u}\left(S_{1}, S_{2}\right)=1-\frac{\left\|S_{1} \otimes S_{2}\right\|_{1}}{\left\|S_{1} \oplus S_{2}-S_{1} \otimes S_{2}\right\|_{1}}
\end{align}
For computational efficiency, we progressively increase the resolutions of silhouettes, obtaining the multiscale mIoU loss $\mathcal{L}_{sp}$, which is represented as:
\begin{align}
\mathcal{L}_{s p}=\sum_{i=1}^{N} \color{black}{\lambda_{si}} \mathcal{L}_{iou}^{i}
\label{lsp}
\end{align}

The predicted viewpoint $\xi_{pred}$ is also used to guide the generation process. We feed the viewpoint into two other fully connected layers $D_v$ to produce a view-aware vector representation $z_v$ and input both $z_v$ and $z_s$ to the decoder $D$ to produce $M_\Theta$. 

A common degradation can occur in which $M_\Theta $ is generated directly from $z_s$ and $z_v$ is completely ignored if the model is trained without any other constraints. To further condition the generation process with the viewpoint constraint, we add a random-view mesh synthesis branch, in which a random viewpoint $\xi_{random}$ is obtained and a mesh $M_{\Theta r}$ is generated in the same manner as mesh generation with $\xi_{pred}$. We use a differentiable renderer to render the silhouettes $S_{\Theta}$ from mesh $M_{\Theta}$ and render the silhouettes $S_r$ from mesh $M_{\Theta r}$. The generated silhouettes $S_{r}$ are regarded as the out-of-distribution fake sample, while the generated silhouettes $S_{\Theta}$ are regarded as the real sample. A shape discriminator $SD$ is introduced to take the inputs of real and fake samples and force the neural network to generate meshes under the view constraint. 
\subsection{3D Model Generation with Structural Awareness}
At this point, the supervision of the mesh generation fidelity is performed with a single rendered silhouette of a generated mesh with a given viewpoint. We find that 2D input alone cannot meet the demand for obtaining complete 3D shapes with \textit{fine-grained structural information} since a single sketch and the corresponding silhouette can only represent the information at that given viewpoint and lacks the information from other viewpoints. Therefore, we propose a random pose sampling (RPS) strategy, which uses multiple random-view silhouettes to supervise the sketch-to-model process. Random pose sampling aims to give the network the capability to generate reasonable 3D fine-structured shapes independent of the viewpoints. As many previous works have investigated in the realm of shape-from-silhouette, the proposed multiview silhouettes contain valuable geometric information about the 3D object~\cite{gadelha2019shape,hu2018structure,zheng2009robust} and thus can serve as effective clues in the 3D model generation process. In addition, during the training process, the Sketch View Prediction Module may encounter degradation, resulting in 3D shapes being generated directly from shape code $Z_s$ and disregarding the significance of viewpoints, consequently impairing its viewpoint awareness. To address this challenge, we introduce the shape discriminator $SD$, which undergoes joint training with the encoder and decoder using an adversarial approach. The integration of random view augmentation during training and the shape discriminator serves to strike a balance between view perception and shape quality. This training strategy mitigates the common degradation issue to a certain extent and enhances the model's viewpoint awareness.

In practice, we randomly sample $N_\xi$ camera poses $\xi_{1...N_\xi}$ from camera pose distribution $ p_{\xi} $. We use a differentiable renderer to render the silhouettes $S_{\Theta}\{1...N_\xi\}$ from the mesh $M_{\Theta}$ and render the silhouettes $S_r\left \{1...N_\xi \right \} $ from the mesh $M_{\Theta r}$. The extrasampled silhouettes of the real mesh and the fake mesh are fed into the discriminator. By introducing $S_r\left \{1...N_\xi \right \} $, the network can use the geometric structure of the objects in cross-view silhouettes while producing the 3D objects, and the discriminator helps to resolve the challenge due to the sparsity of sketches by offering more visual clues. The disentanglement process is very similar to disentangling the ``where" and ``what" principles in generative models \cite{zhu2021and}, which has proven to be effective in our tasks. 

Moreover, the shape discriminator is also carefully designed to fully capture the structural information of the rendered silhouettes. We apply a progressive shape convolutional discriminator $SD$. Following \cite{karras2017progressive}, our discriminator is trained with increasing image resolution and incrementally adds new layers to handle higher resolutions and discriminate fine details. We have found that such a convolutional discriminator design is more effective in capturing local and global structural information to facilitate the generation of high-fidelity 3D shapes compared to the MLP-enabled discriminator for 3D objects. In training, nonsaturating GAN loss with R1 regularization is used \cite{mescheder2018training} for better convergence:

\begin{align}
\begin{split}
\mathcal{L}_{sd} &=\mathbf{E}_{\mathbf{z_v} \sim p_{z_v}, \xi \sim p_{\xi}}\left[f\left(SD_{\theta_{D}}\left(R(M_\Theta, \xi)\right)\right)\right] \\
&+\mathbf{E}_{\mathbf{z_{vr}} \sim p_{z_{vr}}, \xi \sim p_{\xi}}\left[f\left(-SD_{\theta_{D}}(R(M_{\Theta r}, \xi))\right)\right] \label{gan}
\end{split}\\ 
&\textit { where } f(u)=-\log (1+\exp (-u))
\end{align}

\subsection{3D Reconstruction and In-Situ 3D Modeling}
We next apply the sketch-to-model process in a real environment, which is enabled by a state-of-the-art real-time indoor 3D reconstruction algorithm \cite{sun2021neuralrecon} and our customized acquisition application. Specifically, the reconstruction is performed incrementally, with input from the RGB camera and poses. The network directly optimizes the 3D volume represented by a volumetric truncated signed distance function (TSDF) from the inputs, and the mesh is obtained by marching cubes \cite{lorensen1987marching}. Accurate, coherent, and real-time reconstruction can be achieved and displayed via our customized app.

After the surrounding environment is reconstructed, considering a user viewing the mesh of the 3D scene at a specified view in the world coordinates, they can sketch the desired object in that scene immersively. The object belongs to a user-defined class, and the system selects the corresponding weight of the sketch-to-model network based on the class. The sketch is preprocessed and input into the sketch-to-model network. A view estimation of the sketch in the canonical view and a 3D model at that particular view are produced via the sketch-to-model network. A relative position and pose (rotation) can be calculated to place the generated model in the scene at the desired location. Specifically, the rotation is derived from the viewpoint estimation result from the sketch-to-model process, and the translation is derived based on the relative position of the central point within the reconstructed mesh. Algorithm 1 is the pseudocode summarizing the method.

\begin{algorithm}[H]
\caption{Sketch-Based 3D Modeling in a 3D Scene.}\label{alg:alg1}
\begin{algorithmic}
\STATE {\textsc{Begin}}
\STATE \# $M_t$ : mesh of the target scene
\STATE \# $\xi_{t}$ : camera pose of the rendered pictures
\STATE \# $I_t$ : image rendered in a specified camera pose
\STATE \# $I_s$ : freehand sketch masked on $render_{s,c}$
\STATE {\textsc{Input}} $M_t, \xi_{t}, I_t, I_s, class\_id$
\STATE
\STATE \# Sketch preprocessing and use of our method to generate meshes and predict the pose
\STATE $ I_s' \gets resize(crop(I_s))$
\STATE $ M_\Theta, \xi_{pred} \gets Sketch-to-model(I_s')$ 
\STATE
\STATE \# Estimating the relative position and scale of $M_\Theta$ and $M_t$ and converting them into translation and rotation matrices
\STATE $ \bigtriangleup x, \bigtriangleup y, \bigtriangleup z, \bigtriangleup s \gets estimate\_offset(\xi_{t}, I_t, I_s) $
\STATE $ \mathbf{t} \gets compute\_translation(\bigtriangleup x, \bigtriangleup y, \bigtriangleup z) $
\STATE $ \mathbf{R} \gets compute\_rotation(\xi_{pred}) $
\STATE
\STATE \# Transforming $M_\Theta$ and merging it with $mesh_s$
\STATE $ M_\Theta' \gets transform(M_\Theta, \mathbf{R}, \mathbf{t}, \bigtriangleup s) $
\STATE $ M_{f} \gets merge(M_\Theta', M_t) $
\STATE {\textsc{End}}
\end{algorithmic}
\label{alg1}
\end{algorithm}

\section{Experiment}
\subsection{Dataset}
Training the model requires large-scale sketch data with the corresponding 3D models, which are rarely available from publicly accessible sources. Following Zhang et al. \cite{zhang2021sketch2model}, we used the synthetic data ShapeNet-Synthetic for training and testing and the real-world data ShapeNet-Sketch to evaluate the method in the wild. 

ShapeNet-Synthetic is the edge map extracted by a  Canny edge detector from rendered images provided by Kar et al. \cite{kar2017learning}. It contains 13 categories of 3D objects from ShapeNet. ShapeNet-Sketch is a dataset collected from real human drawings. Volunteers with varied drawing skills were asked to draw objects based on the rendered images of 3D objects from Kar's dataset \cite{kar2017learning}, and there are a total of 1300 sketches and their corresponding 3D shapes.

The training of the indoor 3D reconstruction network is based on the commonly used ScanNet-V2 dataset \cite{dai2017scannet}. This dataset is a large-scale resource for indoor 3D scene understanding, containing RGB images, depth images, 3D point cloud data, and semantic and instance annotations from indoor environments.

\subsection{Implementation Details}
For the sketch-to-model process, we utilize ResNet-18 \cite{he2016deep} as the encoder for image feature extraction. The extracted 512-dim feature is processed through two linear layers with L2-normalization, yielding a 512-dim shape code $z s$ and a 512-dim view code $z v$. The rendering module is SoftRas \cite{liu2019soft}, and the number of views is $N=3$. Each 3D object is positioned in the canonical view with a set distance from the camera, 
0 elevation, 
and 0 azimuth angle. We utilize the Adam optimizer with an initial learning rate of 1e-4 that is multiplied by 0.3 every 800 epochs. Beta values are set as 0.9 to 0.999. The total number of training epochs is 2000. 

\input{table/table4}

The loss function $\mathcal{L}$ for the sketch-to-model process is calculated as the weighted sum of five components:
\begin{align}
\mathcal{L} =  \mathcal{L}_{sp} + \mathcal{L}_{r}+ \lambda_v \mathcal{L}_v  + \lambda_{sd} \mathcal{L}_{sd} + \lambda_{dd} \mathcal{L}_{dd} 
\label{loss}
\end{align}
$\mathcal{L}_{r}$ denotes the flattening loss and Laplacian smoothing loss as in~\cite{zhang2021sketch2model, kato2018neural, liu2019soft}, which is used to make the meshes more realistic with higher visual quality.
$\mathcal{L}_{dd} $ is the loss for domain adaptation, as in \cite{zhang2021sketch2model}. The lack of a large amount of ground-truth 3D models and the corresponding 2D sketches leads us to use synthetic data for training and testing on real-world data -- a domain gap exists in the synthetic data and the real-world data. $\mathcal{L}_{dd} $ is thus introduced to make our network generalizable to real hand-drawn datasets. We use domain adaptation on 7 of the classes, which have a sufficient number of sketches in the Sketchy dataset \cite{sangkloy2016sketchy} and Tu-Berlin dataset \cite{eitz2012humans}. Domain adaptation is performed by concatenating the average pooling and max pooling results of the image feature map as input, as in \cite{woo2018cbam}.  $\lambda_{sd}$ and $\lambda_{dd}$ in Equation \ref{loss} equal 0.1, and $\lambda_{v}$ equals 10.

For the 3D reconstruction process, the network was trained following the settings in \cite{sun2021neuralrecon}. To apply the trained network, we wrote a custom Android application that captures videos using the onboard RGB camera of the phone. Along with the captured video, the extrinsic camera information, including the real-time pose, was obtained through the ARCore API. Using the camera poses and the video clips, a key-frame set was selected following the method in \cite{hou2019multi} as the input to the 3D reconstruction network to obtain the predicted mesh of the surrounding environment.

\subsection{Experimental Results for Sketch-View Prediction}
We evaluated the performance of view prediction, which was jointly trained with the sketch-to-model process. We tested the mean absolute error (MAE) of the predicted viewpoint and the ground-truth viewpoint in the ShapeNet-Synthetic dataset, measured in degrees. The result is shown in Table I. Our method achieves state-of-the-art (SOTA) sketch-view prediction performance in elevation and azimuth angles. Note that the azimuth angle has larger errors in some categories (bench, cabinet, display, lamp, loudspeaker, table, telephone), as in these categories, objects have multiple symmetry planes. 

\begin{figure*}[b]
\includegraphics[width=\textwidth]{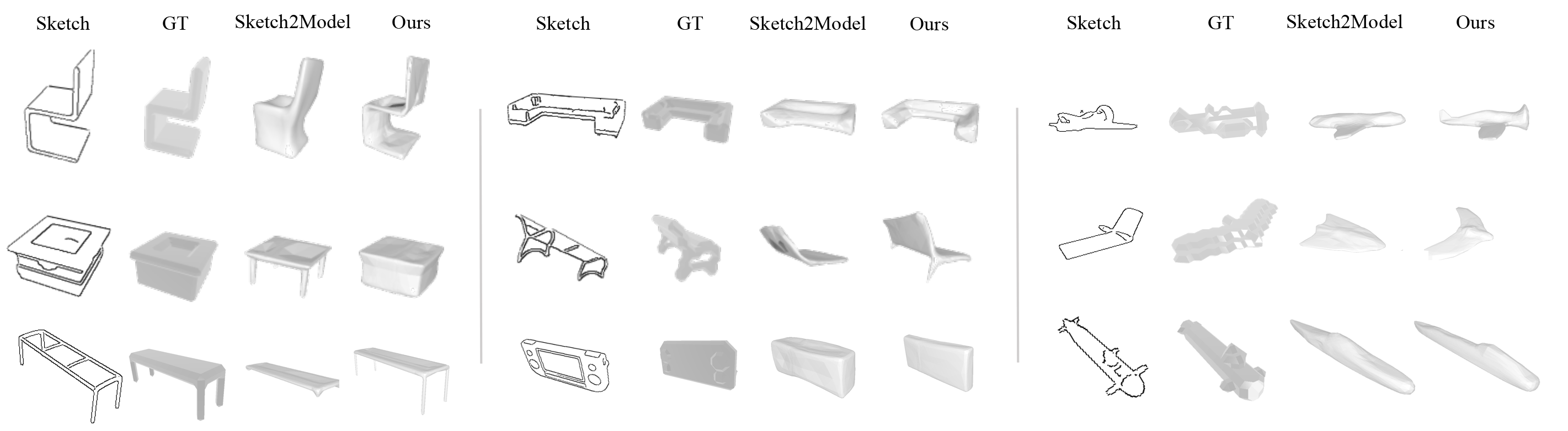}
\caption{{\textbf{Qualitative evaluation with existing state-of-the-art methods}.} The visualization of the generated 3D models demonstrates that our method is capable of synthesizing 3D structures with higher fidelity.}
\end{figure*}

\input{table/table1}

\subsection{Experimental Results for Sketch-to-Model Generation}

\noindent\textbf{The ShapeNet-Synthetic Dataset}

We first evaluated the performance of the dataset with the ground-truth 3D model. Following \cite{zhang2021sketch2model}, we compared our method with a naive autoencoder network, model retrieval with features from a pretrained sketch classification network, and Sketch2Model \cite{zhang2021sketch2model} as the current state-of-the-art (SOTA) model. We first assessed the model's performance using the training/test sets of the \textit{ShapeNet-Synthetic} dataset, which offered precise ground-truth 3D models for training and evaluation purposes. Meshes with the predicted viewpoint (Pred Pos) and the ground-truth viewpoint (GT Pos) were trained and evaluated. We applied a commonly used 3D reconstruction metric -- voxel IoU -- to measure the fidelity of the generated mesh. The results are shown in Table II. The qualitative results demonstrate the effectiveness of our approach with state-of-the-art (SOTA) performance in every category evaluated. 
To verify the statistical significance of this superior performance, we conducted t tests comparing our approach to prior methods. The results confirm that our approach outperforms existing methods with p $<$ 0.05, indicating that the improvements are statistically significant. The quantitative evaluation of our method compared with existing state-of-the-art methods further demonstrated the effectiveness of our approach in reconstructing models with higher structural fidelity, as shown in Figure 4.

\input{table/table2}

\noindent\textbf{The ShapeNet-Sketch Dataset}

We further evaluated the performance on real-world human drawings through the ShapeNet-Sketch dataset. We trained the model on the ShapeNet-Synthetic dataset and used the ShapeNet-Sketch dataset for evaluation. As shown in Table III, our model outperforms the existing state-of-the-art methods in most categories, demonstrating the effectiveness of our approach. In some categories, our method outperforms the existing methods even without domain adaptation (DA). The introduction of DA can further boost the performance in some categories by reducing the gap between real and synthetic data.

\input{table/table3}

After adequately training the network, we tested the neural network on a computer with a graphics card (NVIDIA Tesla V100). Our approach had a generation speed of 123 frames per second (FPS). We also conducted a CPU-only performance test (Intel Xeon E5-2650 V3), and the results showed a 6$\%$ speed boost over Sketch2Model \cite{zhang2021sketch2model} under the same test settings (0.0328 s), with a rate of 30 FPS, which is sufficient for natural computer-human interaction.

\subsection{User Study of the} Immersive 3D Modeling Process 

\begin{figure*}[ht]
\centering\includegraphics[width=0.96\textwidth]{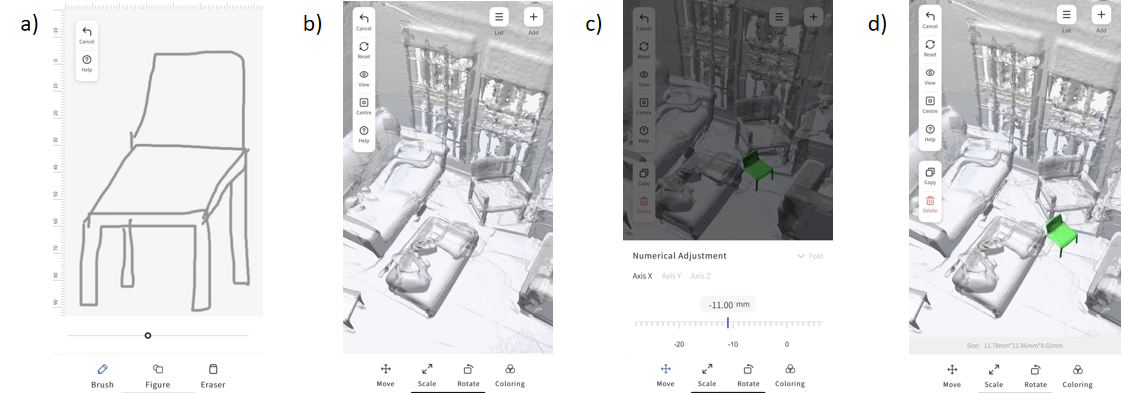}
\caption{{\textbf{Screenshot of the User Study}.} a) A separate blank drawing pad for users to draw sketches. b) The 3D scanned scene shown in the app interface. c) The generated object can be placed and rotated using the built-in function of the app. Users can also use their hands to move the parts. d) Users put the object in their desired place and angle in the scene. (The sketch-derived 3D object is highlighted in green.)}
\label{fig:5}
\end{figure*}

Our immersive 3D modeling experience offers creators the ability to design 3D models that fit the context quickly and efficiently. To validate the effectiveness of our approach, we conducted a user study where we compared the time costs for designers creating models using our approach with sketches drawn over a 3D scanned mesh and a baseline method where designers manually placed a model after designing it separately in a blank drawing pad without context information. The study involved 12 designers with 3D design expertise who drew sketches of chairs on a blank drawing pad (Fig. \ref{fig:5} (a)) and obtained the generated 3D model file. The participants in the study were instructed to use a mobile phone to design a piece of furniture in an office setting. They were given the freedom to adjust the camera angle to find the optimal position for beginning their design. The user interface employed in the study is presented in Fig. \ref{fig:5} and was a custom-designed mobile app that allowed users to draw, place, and view a 3D model of the designed object in situ within the environment. For comparison, we asked the designers to use the mobile app to manually place the generated 3D chair model in the 3D scanned mesh with the built-in "translate," "scale," and "rotate" features (touch-based interaction \cite{martinet2011integrality,goh20193d}), as shown in Fig. \ref{fig:5} (b-d). The total time for designing the 3D models and manually placing them in the context was recorded, and the average time spent using our approach was compared to the average recorded time of the baseline method. The volunteers were asked to perform 3D modeling in each setting 3 times, for a total of 6 times. The results, shown in Table \ref{table:tableVI}, indicate that our method can be more than 5x faster than the baseline method, demonstrating the effectiveness of our approach in enabling rapid and efficient 3D modeling within a scanned context.
\begin{table}[H]
\setlength{\tabcolsep}{6mm}
\caption{The Average Time Comparison of a User Performing 3D Modeling within a Scene}
\begin{center}
\scalebox{1}{
\begin{tabular}{|l|c|}
\hline
             & {Time (s)  $\downarrow$ }  \\ \hline
             
Baseline Method& 121.47 $\pm$ 36.28                                                     \\ \hline
Ours         & \textbf{18.94 $\pm$ 4.28}                                          \\ \hline
\end{tabular}}
\end{center}
\label{table:tableVI}
\end{table}

\subsection{User Study of Geometric Context Information} 
We used further experiments to verify the necessity of introducing geometric context information. Not only could the model learn a new digital environment 3D file with added objects, but this information was also critical for the user's creation process. Specifically, we recruited 12 volunteers and let them use a redesigned user interface, which allowed the users to draw sketches in a captured 2D image. The 3D scene was still reconstructed as the users moved their phones so that the obtained 3D model could remain in a specific 3D position. However, users could only see RGB images, not 3D meshes of the context. We ran a total of 48 sessions. In each session, volunteers were asked to perform the same task (e.g., designing and placing a table next to a sofa) with the two approaches. After completing each modeling task, we asked the volunteers to rotate the camera and view the designed objects from different angles. They were then asked to determine whether a "redo" operation was required due to inaccurate or unrealistic reconstruction or collision issues. We collected the number of "redo" calls for each approach. At the end of all the sessions, we asked the volunteers to evaluate the controllability and usefulness of each approach, which are commonly used criteria for evaluating user interface usability and user experience \cite{albert2022measuring,oh2018lead}. We followed the settings in a prior study \cite{oh2018lead}, using a 7-point Likert scale that ranged from ``highly disagree'' to ``highly agree''. The result in Table \ref{table:table9} shows a higher level of user experience ratings when using the geometric context than the RGB image context.

\begin{table}[ht]
\setlength{\tabcolsep}{4mm}
\caption{The User Experience Evaluation with the 7-point Likert Scale Rating.}
\begin{center}
\scalebox{1}{
\begin{tabular}{| c | c | c |}
\hline
                  & (Q1):Controllability	& (Q2):Usefulness \\ \hline
RGB Image Context & 3.42 ± 0.79	&3.00 ± 0.74 \\ \hline
Geometric Context & 5.17 ± 0.94&	5.08 ± 0.90  \\ \hline
\end{tabular}}
\end{center}
\label{table:table9}
\end{table}

\subsection{Evaluating the Runtime for 3D Modeling}

After adequately training the neural network, we tested it on a computer with an NVIDIA Tesla V100 graphics card. Our approach had a generation speed of 123 FPS. We also conducted a CPU-only performance test (Intel Xeon E5-2650 V3), and the results showed a 6$\%$ speed boost over Sketch2Model \cite{zhang2021sketch2model} under the same test settings (0.0349 s), with a rate of 30 FPS, which is sufficient to be used for natural computer-human interaction.

\begin{table}[ht]
\setlength{\tabcolsep}{8mm}
\caption{Average Runtime for Generating a Single 3D Model from a Sketch}
\begin{center}
\scalebox{1}{
\begin{tabular}{| c | c | c |}
\hline
                  & Speed (s) & FPS \\ \hline
Inference by GPU & 0.0081   & 123 \\ \hline
Inference by CPU & 0.0328   & 30  \\ \hline
\end{tabular}}
\end{center}
\label{table:table6}
\end{table}

\subsection{User Study of 3D Modeling Results}
To further validate the effectiveness of our sketch-to-model algorithm, we conducted a user study following the settings of \cite{cai2021unified,michel2022text2mesh,yao2022dfa} and used the metric of the widely used mean option score (MOS) ranging from 1-5 \cite{seufert2019fundamental} for two factors: Q1: How well does the output 3D model match the input sketch? \textit{(Fidelity)}; Q2: What do you think of the quality of the output 3D model? \textit{(Quality)}. We recruited 12 designers who were familiar with 3D content and presented them with 36 3D modeling results generated by our algorithm. Prior to the experiment, we gave each participant a brief and one-to-one introduction to the concepts of fidelity and quality. We recorded the rating results and averaged the scores. The results are shown in Table \ref{table:tablev}. As perceived by users, our method outperforms existing state-of-the-art methods in the user subject ratings.
\begin{table}[H]
\setlength{\tabcolsep}{6mm}
\caption{Mean Opinion Scores (1-5) for Q1 (Fidelity) and Q2 (Quality)}
\begin{center}
\scalebox{1}{
\begin{tabular}{|l|c|c|}
\hline
             & {(Q1): Fidelity} & {(Q2): Quality} \\ \hline
             
Sketch2Model & 3.36                               & 3.02                              \\ \hline
Ours         & \textbf{3.47}                      & \textbf{3.44}                     \\ \hline
\end{tabular}}
\end{center}
\label{table:tablev}
\end{table}

\subsection{Ablation Study}

\begin{figure}[h]
\includegraphics[width=0.5\textwidth]{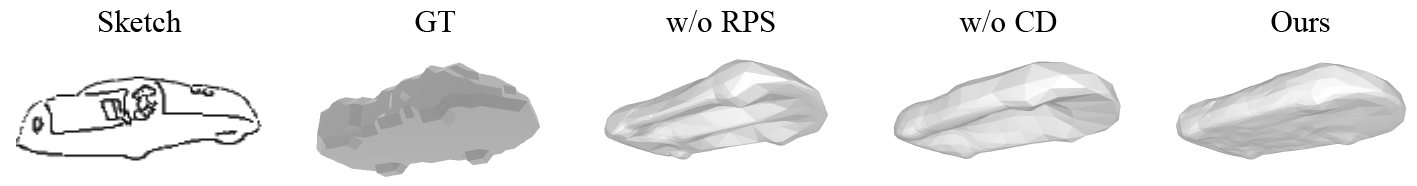}
\caption{\textbf{Visualization of the Ablation Study.} The sketch-to-model network generates unwanted structures w/o RPS and unrealistic structures w/o the SD, showing the effectiveness of RPS and the SD.}
\label{fig:3}
\end{figure}

To show the effectiveness of our proposed method, we conducted an ablation study that removes random pose sampling (RPS) for view awareness. We also removed the progressive shape convolutional discriminator (SD) and used an MLP-based discriminator as in \cite{zhang2021sketch2model}. Our quantitative results (Table \ref{table:tablevi}) and qualitative example (Figure \ref{fig:3}) show that removing the RPS and SD is detrimental to the performance.

\input{table/table6}

Specifically, in RPS, we sampled multiview silhouettes for supervision to generate high-fidelity 3D models. We further performed a sensitivity analysis to determine how the number of sampled views affects the performance of the network. We changed the number of views and trained the neural network again, keeping all the settings and the network structures unchanged. The results are shown in Table VII. From the experimental results, we find that sampling three views brings slightly higher performance than using only two views, which means that multiview images are used to guide the network to facilitate optimization toward higher-fidelity models.

\section{Conclusion}
In this study, we provide a novel solution, \textit{Reality3DSketch}, for 3D modeling. Unlike conventional CAD software, we take advantage of deep neural networks for intuitive and immersive 3D modeling. We demonstrate that users can use their phones to capture the surrounding environment and draw a single-view sketch on the screen. The algorithm reconstructs the 3D mesh of the surrounding environment in real time and produces a 3D object according to the user-drawn sketch \textit{in situ}. We introduce a novel neural network to perform sketch view prediction and 3D modeling with the input of a single sketch. The network is designed to be view- and structure-aware, enabled by random pose sampling (RPS) and a progressive shape discriminator (SD) to produce high-fidelity models. Extensive experiments on both synthetic and real-world datasets demonstrate the effectiveness of our approach. We achieved state-of-the-art (SOTA) performance in both sketch view prediction and 3D modeling. Our user study shows that our method yields $>5$ times faster 3D modeling in a scene compared to separately modeling an object and manually placing it in a scene. Users are also more satisfied with the generated 3D model compared to existing methods. We believe that our work forges a new path and will have great potential to enable creators to perform 3D modeling in the future.  

\section{Limitations and Future Works}
Our current system uses single-view sketches, which inherently lack comprehensive information. Due to this limited input, our 3D shape generation method struggles to produce high-fidelity results when there is heavy occlusion or missing information. With such incomplete input, it is difficult for the network to reliably determine the complete 3D geometry. Future work on incorporating other forms of context could help address these challenges. Currently, the generated 3D scene is only used to assist users in sketching from a single perspective. While this benefits users, the scene information has not yet been utilized to optimize pose estimation or shape generation. Future work could explore leveraging the scene geometry for these purposes. Additionally, since the sketch is in camera coordinates while the context is in world coordinates, investigating how world-space features could inform the model in the camera or canonical space represents another interesting research direction. Overall, our work provides an initial proof of concept, and we believe future research can build on this foundation to enable further applications.

\section*{Acknowledgments}
This paper is supported by the National Key R$\&$D Program of China (2022YFB3303301), National Natural Science Foundation of China (NSFC) (Grant No. 62006208, 62202418), and the National Research Foundation, Singapore under its AI Singapore Programme (AISG Award No: AISG2-PhD-2021-08-006). Tianrun Chen acknowledges funding from KOKONI, Moxin (Huzhou) Technology Co., LTD and Moxin Technology (HK). The author thanks Papa Mao and Xin Xu for discussion.

\bibliographystyle{IEEEtran}
\bibliography{strings}

\begin{IEEEbiography}[{\includegraphics[width=1in,height=1.25in,clip,keepaspectratio]{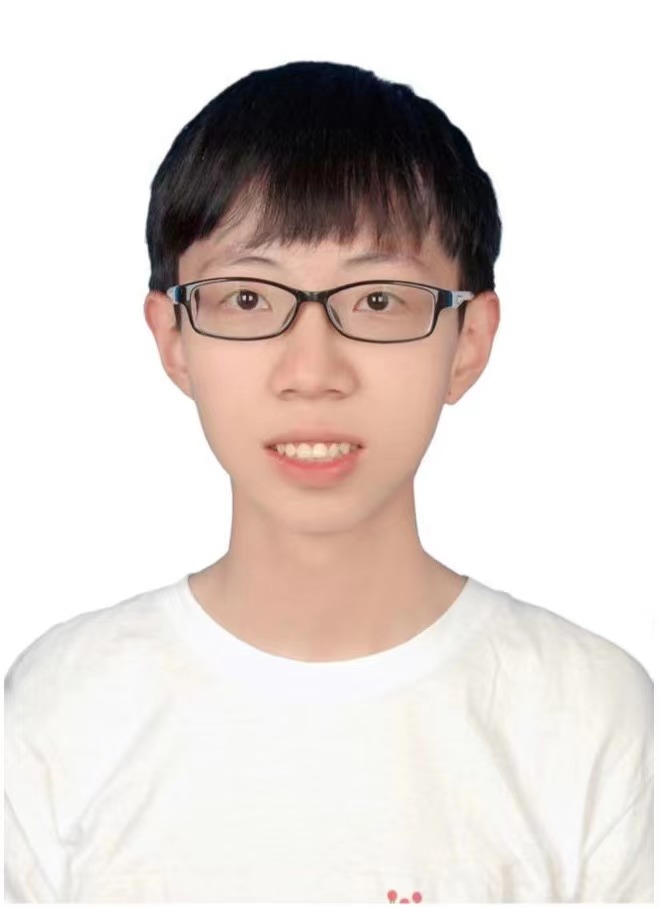}}]{Tianrun Chen}
received a bachelor's degree from the College of Information Science and Electronic Engineering, Zhejiang University, and is pursuing a Ph.D. degree at the College of Computer Science and Technology, Zhejiang University. He is the founder and technical director of Moxin (Huzhou) Technology Co., LTD. His research interests include computer vision and its enabling applications. 
\end{IEEEbiography}
\begin{IEEEbiography}[{\includegraphics[width=1in,height=1.25in,clip,keepaspectratio]{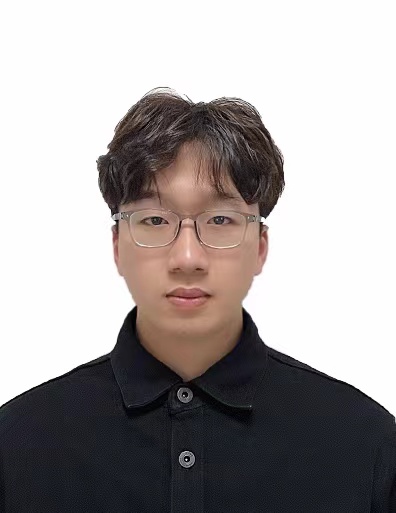}}]{Chaotao Ding}

is currently studying for a master's degree in electronic information at Huzhou University, focusing on 3D reconstruction and computer vision. He is a student member of CCF, and he has published articles in several computer vision-related journals.
\end{IEEEbiography}
\begin{IEEEbiography}[{\includegraphics[width=1in,height=1.25in,clip,keepaspectratio]{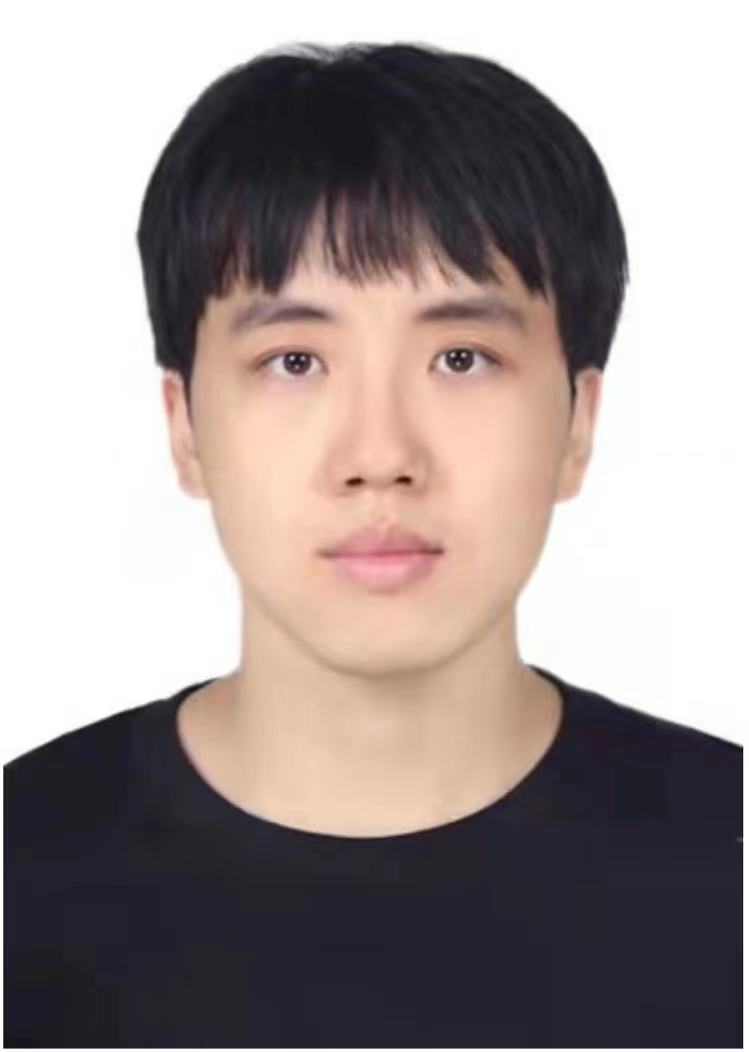}}]{Lanyun Zhu}
received his B.E. degree from Beihang University, Beijing, China in 2020. He is currently pursuing a Ph.D. degree with the Information Systems Technology and Design (ISTD) pillar, Singapore University of Technology and Design. His research interests are mainly focused on deep learning and computer vision. He is the reviewer of multiple top journals and conferences, including IEEE T-IP, ICML and NeurIPS.
\end{IEEEbiography}
\begin{IEEEbiography}[{\includegraphics[width=1in,height=1.25in,clip,keepaspectratio]{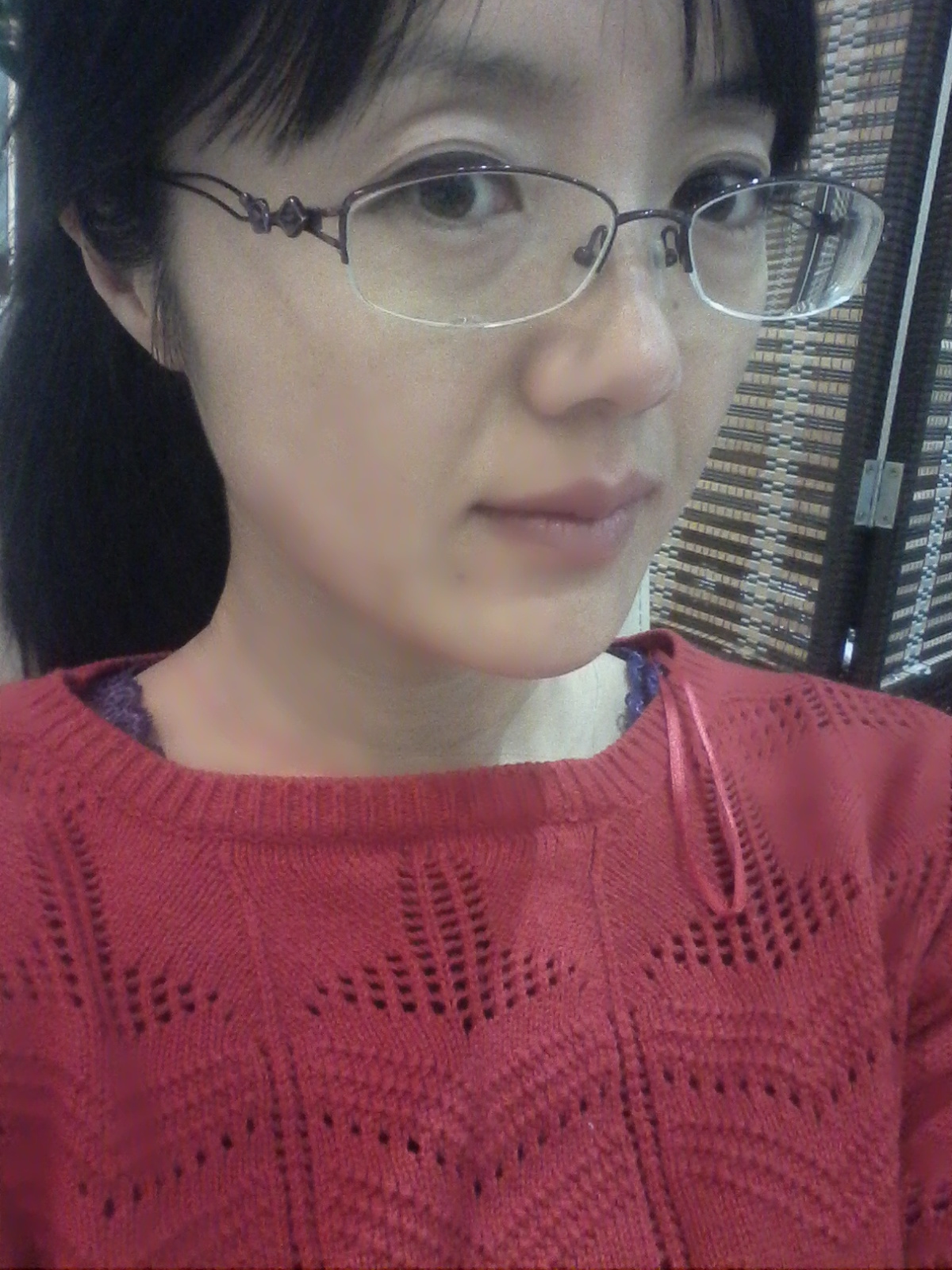}}]{Ying Zang} received her B.S. degree in computer science and technology from Liaoning University, China, in 2004; her M.S. degree in computer science and technology from Dalian Maritime University, China, in 2010; and her Ph.D. degree in computer application technology from Chinese Academy of Sciences University, China, in 2022. She is an AI engineer at the School of Information Engineering of Huzhou University. She is currently working on research on 3D vision, object detection and semantic segmentation.
\end{IEEEbiography}
\begin{IEEEbiography}[{\includegraphics[width=1in,height=1.25in,clip,keepaspectratio]{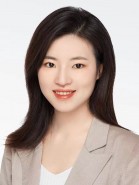}}]{Yiyi Liao}
received her Ph.D. degree from the College of Control Science and Engineering, Zhejiang University, China, in 2018. She is currently an assistant professor at the College of Information Science and Electronic Engineering, Zhejiang University. Her research interests include 3D vision and scene understanding.
\end{IEEEbiography}
\begin{IEEEbiography}[{\includegraphics[width=1in,height=1.25in,clip,keepaspectratio]{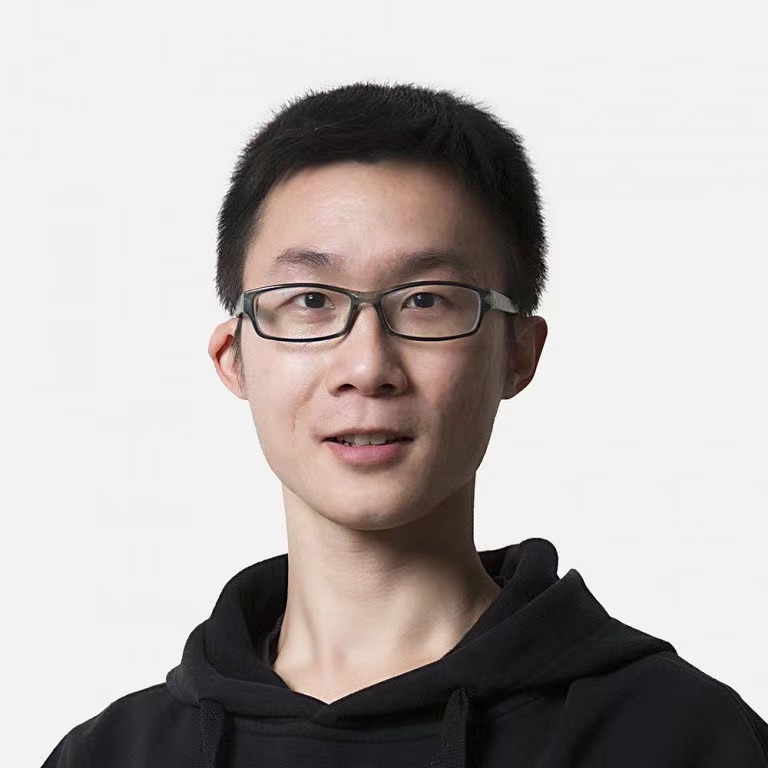}}]{Zejian Li}
is an assistant researcher at the School of Software Technology, Zhejiang University. He obtained a Ph.D. degree from Zhejiang University. His research interests include generative models, interpretable image generation and intelligent design.
\end{IEEEbiography}
\begin{IEEEbiography}[{\includegraphics[width=1in,height=1.25in,clip,keepaspectratio]{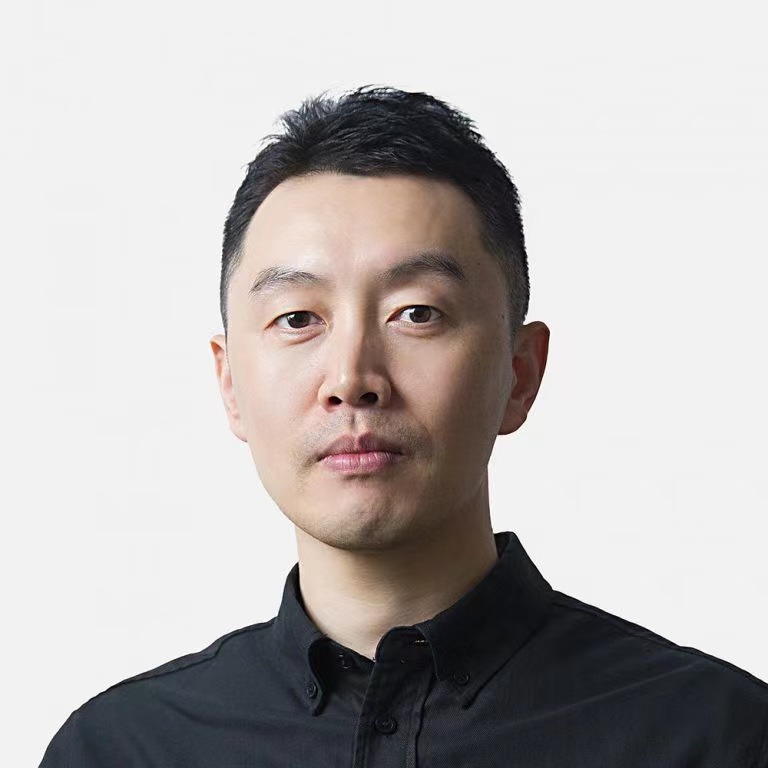}}]{Lingyun Sun}
is a professor at the School of Computer Science and Technology, Zhejiang University. He obtained a Ph.D. degree from Zhejiang University.
His research revolves around AI and design, aiming to equip the design industry with AI capabilities and to enhance design tools and methodologies in the AI era. He has developed image and video generation platforms that can create visual content, short videos, and other digital materials.
\end{IEEEbiography}
\end{document}

%% file: table/table4.tex
\begin{table*}[h]
\begin{center}
\caption{{The quantitative evaluation of sketch-view estimation}}
\scalebox{1}{
\begin{tabular}{|c|c|c|c|c|c|c|c|c|}
\hline
\multicolumn{9}{|c|}{Shapenet-synthetic (MAE $\downarrow)$} \\
\hline
\multicolumn{2}{|c|}{}  & car & sofa & airplane & bench & display & chair & table \\
\hline
\multirow{2}{*}{Elevation} & Sketch2Model & 1.0751 & 1.5989 & 2.3899 & 1.8345 & 1.8944 & 1.8690 & 1.2857  \\
\cline{2-9} & \textbf{Ours} & \textbf{0.9029} & \textbf{1.395} & \textbf{2.2014} & \textbf{1.0168} & \textbf{1.6826} & \textbf{1.5422} & \textbf{1.0184} \\
\hline
\multirow{2}{*}{Azimuth} & Sketch2Model & 5.0986 & 11.0327 & 10.4171 & 43.7923 & 44.1861 & 8.6753 & \textbf{86.7654}  \\
\cline{2-9} & \textbf{Ours} & \textbf{4.3056} & \textbf{9.8532} & \textbf{9.7180} & \textbf{38.7755} & \textbf{43.2417} & \textbf{7.2630} & 87.6369 
 \\
\hline
\multicolumn{2}{|c|}{} & telephone & cabinet & loudspeaker & watercraft & lamp & rifile & mean \\
\hline
\multirow{2}{*}{Elevation} & Sketch2Model & 2.2732 & 1.2148 & 2.4303 & 3.6884 & 4.4071 & 3.3226 & 2.2526 \\
\cline{2-9} & \textbf{Ours} & \textbf{2.0720} & \textbf{1.0168} & \textbf{2.0659} & \textbf{3.4014} & \textbf{4.0796} & \textbf{3.1199} & \textbf{1.9627} \\
\hline
\multirow{2}{*}{Azimuth} & Sketch2Model & 54.3659 & 41.7126 & 73.8672 & 34.5512 & \textbf{84.4146} & 11.2999 & 39.2445  \\
\cline{2-9} & \textbf{Ours} &  \textbf{51.2568} & \textbf{38.7755} & \textbf{72.6029} & \textbf{33.8180} & 84.7734 & \textbf{10.7342} & \textbf{37.9042} \\
\hline
\end{tabular}}
\end{center}

\label{table:table1}

\end{table*}




%% file: table/table1.tex
\begin{table*}[ht]
\centering
\caption{The quantitative evaluation of ShapeNet-Synthetic dataset}
\scalebox{0.88}{  
\begin{tabular}{|c|c|c|c|c|c|c|c|}
\hline
\multicolumn{8}{|c|}{Shapenet-synthetic (Voxel IoU $\uparrow$)} \\
\hline
 & car & sofa & airplane & bench & display & chair & table \\
\hline
Retrieval & 0.667 & 0.483 & 0.513 & 0.38  & 0.385 & 0.346 & 0.311 \\
\hline
Auto-Encoder & 0.769 & 0.613 & 0.576 & 0.467 & 0.541 & 0.496 & 0.512 \\
\hline
Sketch2Model (GT Pos)  & 0.751 $\pm$ 0.119 & 0.622 $\pm$ 0.163 & 0.624 $\pm$ 0.144 & 0.481 $\pm$ 0.149 & 0.604 $\pm$ 0.135 & 0.522 $\pm$ 0.154 & 0.478 $\pm$ 0.181 \\
Sketch2Model (Pred Pos) & 0.746 $\pm$ 0.124 & 0.620 $\pm$ 0.161 & 0.618 $\pm$ 0.149  & 0.477 $\pm$ 0.153 & 0.550 $\pm$ 0.134 & 0.515 $\pm$ 0.155 & 0.470 $\pm$ 0.181 \\
\hline
\textbf{Ours (GT Pos)} & \textbf{0.796 $\pm$ 0.132} & \textbf{0.651 $\pm$ 0.159} & \textbf{0.644 $\pm$ 0.149} & \textbf{0.500 $\pm$ 0.153} & \textbf{0.612 $\pm$ 0.182}	& \textbf{0.544 $\pm$ 0.152} & \textbf{0.518 $\pm$ 0.187} \\
\textbf{Ours (Pred Pos)} & 0.793 $\pm$ 0.133 & 0.649 $\pm$ 0.158 & 0.641 $\pm$ 0.153 & \textbf{0.500} $\pm$ 0.153 & 0.583 $\pm$ 0.195 & 0.541 $\pm$ 0.152 & 0.504 $\pm$ 0.190 \\
\hline
 & telephone & cabinet & loudspeaker & watercraft & lamp & rifle & mean \\
\hline
Retrieval & 0.622 & 0.518 & 0.468 & 0.422 & 0.325 & 0.475 & 0.455\\
\hline
Auto-Encoder & 0.706 & 0.663 & 0.629 & 0.556 & 0.431 & 0.605 & 0.582 \\
\hline
Sketch2Model (GT Pos)  & 0.719 $\pm$ 0.207 & 0.701 $\pm$ 0.209 & 0.641 $\pm$ 0.208 & 0.586 $\pm$ 0.161 & 0.472 $\pm$ 0.208 & 0.612 $\pm$ 0.170 & 0.601\\
Sketch2Model (Pred Pos) & 0.673 $\pm$ 0.216 & 0.667 $\pm$ 0.220 & 0.624 $\pm$ 0.216 & 0.569 $\pm$ 0.168 & 0.463 $\pm$ 0.209 & 0.606 $\pm$ 0.172 & 0.584 \\
\hline
\textbf{Ours (GT Pos)} & \textbf{0.738 $\pm$ 0.201} & \textbf{0.705 $\pm$ 0.209} & \textbf{0.651 $\pm$ 0.206} & \textbf{0.595 $\pm$ 0.158} & \textbf{0.469 $\pm$ 0.207} & \textbf{0.619 $\pm$ 0.174} & \textbf{0.618}  \\
\textbf{Ours (Pred Pos)} & 0.680 $\pm$ 0.232 & 0.683 $\pm$ 0.216 & 0.623 $\pm$ 0.212 & 0.580 $\pm$ 0.164 & 0.465 $\pm$ 0.209 & \textbf{0.619 $\pm$ 0.174} & 0.604\\
\hline
\end{tabular}
}
\end{table*}

%% file: table/table2.tex

\begin{table*}[ht]
\begin{center}
\caption{The quantitative evaluation of ShapeNet-Sketch dataset}
\scalebox{0.84}{

\begin{tabular}{|c|c|c|c|c|c|c|c|c|c|c|c|c|c|c|c|}

\hline
\multicolumn{15}{|c|}{Shapenet-sketch (Voxel IoU $\uparrow$)} \\
\hline
 & car & sofa & airplane & bench & display & chair & table & telephone & cabinet & loudspeaker & watercraft & lamp & rifile & mean\\
\hline
Retrieval & 0.626 & 0.431 & 0.411 & 0.219 & 0.338 & 0.238 & 0.232 & 0.536 & 0.431 & 0.365 & 0.369 & 0.223 & 0.413 & 0.370\\
\hline
Auto-Encoder & 0.648 & 0.534 & 0.469 & 0.347 & 0.472 & 0.361 & 0.359 & 0.537 & 0.534 & 0.533 & 0.456 & 0.328 & 0.541 & 0.372 \\
\hline
Sketch2Model (GT Pos)  & 0.659 & 0.534 & 0.487 & 0.366 & 0.479 & 0.393 & 0.357 & 0.554 & \textbf{0.568} & 0.526 & 0.450 & \textbf{0.338} & 0.534 & 0.483 \\
Sketch2Model (Pred Pos) & 0.649 & 0.528 & 0.479 & 0.357 & 0.435 & 0.383 & 0.361 & 0.551 & 0.547 & 0.544 & 0.466 & 0.336 & 0.510 & 0.470\\
\hline
Sketch2Model + DA (GT Pos) & 0.679  & \textbf{0.548} & \textbf{0.526} & \textbf{0.367} & - & \textbf{0.398} & 0.357 & - & - & - & - & - & 0.535 & 0.489\\
Sketch2Model + DA (Pred Pos) & 0.659 & 0.533 & 0.515 & 0.362 & - & 0.385 & 0.360  & - & - & - & - & - & 0.511 & 0.475 \\
\hline
\textbf{Ours (GT Pos)} & 0.695 & 0.528 & 0.502 & 0.364 & \textbf{0.493} & 0.389 & 0.370 & \textbf{0.574} & 0.563 & \textbf{0.538} & \textbf{0.477} & 0.334 & 0.535 & 0.489 \\
\textbf{Ours (Pred Pos)} & 0.683 & 0.523 & 0.502 & 0.364 & 0.493 & 0.389 & \textbf{0.370} & 0.527 & 0.549 & 0.509 & 0.468 & 0.331 & 0.535 & 0.476 \\
\hline
\textbf{Ours + DA (GT Pos)}   & \textbf{0.699} & 0.538 & 0.517 & 0.362 & - & 0.390 & 0.360 & - & - & - & - & - & \textbf{0.545} & \textbf{0.491} \\
\textbf{Ours + DA (Pred Pos)} & 0.692 & 0.532 & 0.515 & 0.360 & - & 0.382 & 0.346  & - &  & - & - & - & \textbf{0.545} & 0.477\\
\hline

\end{tabular}}
\end{center}

\label{table:table3}

\end{table*}

%% file: table/table3.tex
\begin{table*}[h]
\caption{{The quantitative evaluation of ablation study.} }
\begin{center}
\scalebox{1}{
\begin{tabular}{|c c | c| c| c| c| c| c| c |}
\hline
\multicolumn{9}{|c|}{Ablation Study. (Numbers inside and outside the parenthesis are IoU on Pred View and GT View, respectively)} \\
\hline
RPS & {\color{black}SD} & car & sofa & airplane & bench & display & chair & table \\
\hline
 & & 0.747 (0.753) & 0.624 (0.643) & 0.557 (0.565) & 0.345 (0.460) & 0.457 (0.577) & 0.499 (0.508) & 0.406 (0.427) \\
 \hline
$\surd$ & & 0.782 (0.773) & 0.641 (0.639) & \textbf{0.644} (0.639) & 0.461 (0.485) & 0.597 (0.540) & 0.543 (0.538) & 0.512 (0.477) \\
\hline
$\surd$ & $\surd$ & \textbf{0.796} (\textbf{0.793}) & \textbf{0.651} (\textbf{0.649}) & \textbf{0.644} (\textbf{0.641}) & \textbf{0.500} (\textbf{0.500}) & \textbf{0.612} (\textbf{0.583}) & \textbf{0.544} (\textbf{0.541}) & \textbf{0.518} (\textbf{0.504}) \\
\hline
RPS & {\color{black}SD} & telephone & cabinet & loudspeaker & watercraft & lamp & rifile & mean \\
\hline
 & & 0.522 (\textbf{0.705}) & 0.597 (0.579) & 0.584 (0.614) & 0.574 (0.575) & 0.290 (0.421) & 0.500 (0.576) & 0.516 (0.569) \\
 \hline
$\surd$ & & 0.734 (0.673) & 0.696 (0.645) & 0.636 (0.599) & 0.585 (0.553) & \textbf{0.478} (\textbf{0.471}) & \textbf{0.619} (\textbf{0.627}) & 0.608 (0.588) \\
\hline
$\surd$ & $\surd$ & \textbf{0.738} (0.680) & \textbf{0.705} (\textbf{0.683}) & \textbf{0.651} (\textbf{0.623}) & \textbf{0.595} (\textbf{0.580}) & 0.469 (0.465) & \textbf{0.619} (0.619) & \textbf{0.618} (\textbf{0.604}) \\
\hline
\end{tabular}}
\end{center}

\label{table:tablevi}

\end{table*}

%% file: table/table6.tex





\begin{table*}[h]
\begin{center}
\caption{The sensitivity analysis for sampling different number of silhouettes}
\scalebox{0.84}{
\begin{tabular}{|c|c|c|c|c|c|c|c|c|c|c|c|c|c|c|}
\hline
\multicolumn{15}{|c|}{Shapenet-synthetic (Voxel IoU $\uparrow)$} \\
\hline
 & car & sofa & airplane & bench & display & chair & table & telephone & cabinet & loudspeaker & watercraft & lamp & rifile & mean \\
\hline

2 Random Views (GT Pos)  & 0.795 & 0.641 & 0.642 & \textbf{0.502} & 0.610 & 0.538 & 0.494 & 0.723 & 0.697 & \textbf{0.658} & 0.586 & \textbf{0.470 }& \textbf{0.624} & 0.617 \\
2 Random Views (Pred Pos) & 0.792 & 0.640 & 0.639 & \textbf{0.502} & 0.592 & 0.535 & 0.489 & 0.675 & 0.676 & 0.638 & 0.582  & 0.465 & \textbf{0.624 }& 0.603 \\
\hline
3 Random Views (GT Pos) & \textbf{0.796} & \textbf{0.651} &\textbf{0.644} & 0.500 & \textbf{0.612}	& \textbf{0.544} & \textbf{0.518} & \textbf{0.738 }&\textbf{0.705} & 0.651 & \textbf{0.595 }& 0.469 & 0.619 &\textbf{0.618} \\
3 Random Views (Pred Pos) & 0.793 & 0.649 & 0.641 & 0.500 & 0.583 & 0.541 & 0.504  & 0.680 & 0.683 & 0.623 & 0.580 & 0.465 & 0.619 & 0.604 \\
\hline

\end{tabular}}
\end{center}

\label{table:tablevii}

\end{table*}